\documentstyle[preprint,aps,floats,epsfig]{revtex}
\tighten

\begin{document}
\draft
\preprint{\ 
\begin{tabular}{rr}
& 
\end{tabular}
} 
\title{Polarization-Temperature Correlation from a Primordial Magnetic Field 
}
\author{Evan S. Scannapieco and Pedro G. Ferreira  }
\address{Center for Particle Astrophysics,  and  Departments of Astronomy and Physics,\\
University of California, Berkeley CA 94720-7304.}
\maketitle

\begin{abstract}
We propose a new method for constraining a primordial
homogeneous magnetic field with the cosmic microwave background. 
Such a field will induce an observable parity odd cross
correlation between the polarization
anisotropy and the temperature anisotropy by Faraday rotation.
We analyze the necessary experimental features to match, and improve, 
current constraints of such a field by measuring this correlation.
\end{abstract}
\date{\today}
\renewcommand{\thefootnote}{\arabic{footnote}} \setcounter{footnote}{0}

In our Galaxy, as well as other spirals, large scale magnetic fields 
on the order of $10^{-6}$ Gauss have been observed, the origin of which 
is unknown \cite{a}.  Schemes that invoke a dynamo mechanism to explain these
fields all rely on the presence of an initial seed field \cite{b},
prompting numerous explanations ranging from nonlinear battery mechanisms,
to changes in the nature of the electroweak force \cite{f}.
Furthermore, the mechanism of dynamo generation itself has yet to be
developed in a fully self-consistent manner, and the maximum possible
fluxes may have been overestimated \cite{c}.
Related to the origin of magnetic fields in spiral galaxies 
is the question of the origin of the $10^{-6}$G fields
detected in high redshift galaxies \cite{d} and in the 
damped Ly$\alpha$ clouds \cite{e}.  At these early times
a plausible dynamo explanation has yet to be proposed.
Alternatively, large scale galactic and extragalatic fields can be 
explained by the adiabatic compression of a primordial magnetic field 
on the order of $10^{-9}$ Gauss today \cite{thorne}.  
Thus, while it is certainly
possible that several of the physical schemes being studied may
play a role in the formation of large scale magnetic fields, the considerable
debate surrounding this subject can not be resolved
until the initial magnetic field configuration is better known. 
This can only come from further observation.

The cosmic microwave background (CMB) supplies us with the
oldest and most extensive probe of the early universe. It has
recently been argued that it may provide tight constraints
on primordial magnetic fields.  In \cite {adams} it was shown that
a random magnetic field with an amplitude of order $10^{-8}$G
would lead to distortions in the angular power spectrum of the CMB
of around $10\%$, observable with planned satellite experiments.
In \cite{BFS} it was shown that, if one included the full anisotropic
effects of the magnetic field one could use the existing large angle
measurements of the COBE satellite to set a limit on a homogeneous 
primordial magnetic field today, 
${\cal B}_0 < 6.8 \times 10^{-9} (\Omega_0 h^2)^{1/2}$ G
where $\Omega_0 \le 1$ is the cosmological density parameter and
the Hubble constant is
$H_0=100h$ km s$^{-1}$ Mpc$^{-1}$. Other authors have suggested that
Faraday rotation might affect the CMB anisotropy and polarization in
a distinct way. In \cite{milan}, the authors derived analytic
expressions for its effect on the dipole, quadropole, and octopole
of the CMB anisotropy in a Bianchi I cosmology.
More recently in \cite{hhz}, the authors found the effect that a 
homogeneous field with
an amplitude of $10^{-8}$G would have in decreasing the polarization
of the CMB, while in \cite{KL}
it was proposed 
that a two frequency measurement of the polarization 
should be able to set bounds on a random magnetic field on the
order of $10^{-9}$G.

In the case of a homogeneous magnetic field  
a more direct single frequency measurement is possible.
The basis of this technique is to exploit the parity and symmetry properties
that such a field induces in the CMB.  Given a measurement of
the Stokes parameters, as a function of position on the sky, this
data can be decomposed into a sum of generalized spherical harmonics.
Thus 
\begin{eqnarray}
I({\hat{\bf n}}) & = & 
\sum_{lm} a^T_{lm} Y_{lm}({\hat{\bf n}}) ; \nonumber \\
(Q \pm iU)({\hat{\bf n}}) & = & 
\sum_{lm} a^{\pm 2}_{lm} {_{\pm 2}}Y_{lm}({\hat{\bf n}})
\end{eqnarray} 
where $T$ denotes temperature and the spin $\pm 2$ spherical
harmonics, $_{\pm 2}Y_{lm}$, are used to preserve the 
behavior of $Q$ and $I$ under coordinate
rotations.  Here we follow the notation discussed in  \cite{h}; for 
an alternative, but equivalent approach see \cite{h2}.  
Circular polarization can be decomposed as
a scalar, but is not expected to occur in the CMB \cite{i}.

Under parity inversion $_sY_{\ell m} \rightarrow (-1)^\ell 
_{-s}Y_{\ell m}$ and thus $_2Y_{\ell m} \pm _{-2}Y_{\ell m}$ 
are parity eigenstates.
This motivates us to define
\begin{eqnarray}
2 a^E_{\ell m} \equiv -(a^{2}_{\ell m}+ a^{-2}_{\ell m}) \qquad {\rm and} \qquad
2 a^B_{\ell m} \equiv i(a^{2}_{\ell m}- a^{-2}_{\ell m}) \label{EBdef}
\end{eqnarray}
so that $a^E_{\ell m} \rightarrow (-1)^\ell a^E_{\ell m}$ and
$a^B_{\ell m} \rightarrow (-1)^\ell a^B_{\ell m}$ under parity inversion.
In an isotropic universe, cross correlations between the B polarization
and the temperature and E polarizations are forbidden as this would imply
noninvariance under parity.  A homogeneous magnetic field, however, is
maximally parity violating and therefore one would expect one of 
its primary signals to be such a cross correlation.   
Motivated by this expectation, we consider in some detail
the effect of a homogeneous magnetic field on CMB polarization.

Polarization of the CMB is generated in the optically thin last scattering 
surface by quadropole fluctuations in temperature.  
Scalar fluctuations result in polarization that is purely of the E sort,
while tensor fluctuations result in B and E polarization.  The
presence of a magnetic field, however, will create B polarization that is
correlated with the measured temperature spectra.  This happens as the 
the B and E polarizations are Faraday rotated into each other.
Following \cite{KL}, we can estimate the extent of this coupling
by taking the optical depth of the last scattering surface to be 
approximately 1.  For a homogeneous magnetic field we find 
\begin{eqnarray}
\phi \approx 2.3^\circ 
\left( \frac{{\cal B}_0}{10^{-9} {\rm G}}  \right) 
\left(\frac{\nu_0}{{\rm 30GHz}}\right)^{-2}\cos \theta
\end{eqnarray}
where $\theta$ is the angle of the magnetic field with respect to the direction
of propagation.  This effect 
can be used to determine the magnetic field by comparing observations
at different frequencies \cite{KL}.
As this B-polarization signal is expected to be small, however, one would
like to minimize experimental noise as well as systematic uncertainties
by comparing it to quantities at the same frequency and with greater 
amplitude.  Given our discussion above $\langle a^{T*}_{\ell',m'} a^B_{\ell,m} 
\rangle$
 represents just such a comparison.  We would therefore like
to examine this quantity in greater detail.

The evolution of radiation in a perturbed Friedman-Robertson-Walker
universe can be studied in the linear regime.  In this case
each plane wave perturbation can be considered separately, with an average
over the ${\bf {\vec k}}$'s being done at the end of the calculation.  
Defining $\mu={{\bf \hat n}}\cdot{{\bf \hat k}}$ where ${\bf{\hat n}}$
is the line
of sight we can write down the Boltzmann
equation in the synchronous gauge as follows \cite{MB}:
\begin{eqnarray}
\dot\Delta_T + ik\mu \Delta_T
=&-&{1\over 6}\dot h-{1\over 6}(\dot h+6\dot\eta)
P_2(\mu) \nonumber \\
&+&\dot\kappa\left[-\Delta_T +
\Delta_{T0} +i\mu v_b +{1\over 2}P_2(\mu)\Pi
\right] \label{ev1} \\
\dot\Delta_Q +ik\mu \Delta_Q =& &\dot\kappa \left[
-\Delta_Q +
{1\over 2} [1-Q_2(\mu)] \Pi\right]  + 2 \omega_B \Delta_U \label{ev2}\\
\dot\Delta_U +ik\mu \Delta_U =& -&\dot\kappa 
\Delta_U -2 \omega_B \Delta_Q \label{ev3} \\
\Pi=& &\Delta_{T2}
+\Delta_{Q2} +
\Delta_{Q0}
\nonumber
\end{eqnarray}
where $\Delta_T$, $\Delta_Q$, $\Delta_U$ are the brightness functions
for $T$, $Q$ and $U$, which can be expanded in  multipole moments
defined such that $\Delta(\tau,k,\mu) = \sum_\ell (2\ell+1) 
(-i)^\ell \Delta_\ell(\tau,k) P_\ell(\mu)$
($P_\ell(\mu)$ is the Legendre polynomial of order $\ell$). Derivatives are 
taken with respect to the conformal time $\tau$, 
$\dot \kappa$ is the differential optical depth for Thomson scattering
$\dot \kappa = (a/a_0) n_e \sigma_T$ and $h$,
$\eta$, $v_b$ are sources of temperature anisotropies from metric
perturbations and baryon velocities. We shall restrict ourselves
to scalar perturbations. The effect of Faraday rotation comes
in through the coupling between polarization components, $w_B={\dot
\kappa}\frac{2}{3}\omega_0\cos\theta$  where
\begin{eqnarray}
\omega_0 = .06
\left( \frac{{\cal B}_0}{10^{-9} {\rm G}} \right) 
\left(\frac{\nu_0}{{\rm 30GHz}}\right)^{-2}
\end{eqnarray}
and we have assumed that ${\cal B}_0$ lies along the z axis.
From Eq. \ref{EBdef} we see that $\Delta_Q$ ($\Delta_U$) give us the
amplitude of the E (B) component of the polarization.

Before we proceed, we shall now discuss two important approximations.
Firstly we will consider an {\it isotropic} universe supporting
a homogeneous magnetic field. As argued in \cite{BFS} this is
not entirely correct, such a field can only be supported by
the existence of a globally anisotropic term. However, as we shall
see, most of our statistic will rely on the large $\ell$ behavior
of the cross correlation spectrum and will therefore be insensitive
to the large scale behavior of the anisotropy. Clearly a more
detailed calculation should include such terms (for a good 
description see \cite{BT}) but for the purpose of this letter we
shall not pursue it. Secondly, we are interested in studying ${\cal
B}_0$ in the regime where it is competitive with other constraints
from the CMB, notably \cite{BFS}, and consequently $\omega_B$ is
sufficiently small that we can drop it from Eq. \ref{ev2}. This
means that the only source term for the correlated B polarization
is through a rotation of E polarized light by $\sim 2^\circ$; 
it is expected to be small compared to E and the coupling of B 
into E even smaller. Thus the effect can be applied only to the 
evolution of B polarized light, and there only as a scale factor of 
${\cal B}_0 \cos \theta$ in the source term.

This represents a great simplification in the calculation.
As the angle of the magnetic field with respect to the
observer appears only as a scale factor, we can sidestep the issue
of the relative orientation of the wave vector ${\bf {\hat k}}$ and
${\vec {\cal B}}_0$, 
using it only in the angular averages taken at the end of the
calculation.  With this simplification we find the cross
correlation due to scalar perturbations to be
\begin{equation}
\langle a_{\ell^\prime m^\prime}^{T*} a^B_{\ell m}\rangle =    
C^{TB}_{\ell',\ell}
(\hat{\bf{z}})
\left[ \int d\Omega Y^*_{\ell'm'}(\hat{{\bf n}}) 
Y_{\ell m}(\hat{{\bf n}}) \cos \theta 
\right] 
\label{eq:crosscor}
\end{equation}
where 
\begin{eqnarray}
 C^{TB}_{\ell',\ell}(\hat{\bf{z}}) & \equiv &
\frac{3 \omega_0}{4} (4 \pi)^2 
\left({(\ell+2)!\over (\ell-2)!}\right)^{1\over 2}
\Big[\int k^2 dk P(k) 
 \int_0^{\tau_0} d\tau g(\tau)\nonumber \\  
 \int_0^{\tau} d\tau'& g(\tau')
 &\frac{j_\ell(x')}{x'^2}
\Pi(k,\tau') 
\int_0^{\tau_0} d\tau'' S_T(k,\tau'') \; j_{\ell'}(x'')
\Big]\nonumber \\
S_T(k,\tau)&=&g(\Delta_{T0}+2{\dot \alpha}+\frac{{\dot v}_b}{k}
+\frac{\Pi}{4}+\frac{3\Pi}{4k^2})\nonumber \\
& &+e^{-\kappa}({\dot \eta}+{\ddot \alpha})+{\dot g}(\frac{v_b}{k}
+\alpha+\frac{3\Pi}{4k^2})+\frac{3{\ddot g}\Pi}{4k^2}\nonumber \\
& &\label{eq:ctb}
\end{eqnarray}
$P(k)$ is the initial power spectrum of
scalar perturbations,  $x \equiv k (\tau_0 - \tau)$, $\alpha=({\dot
h}-6{\dot \eta})/2k^2$, 
$j_\ell(x)$ is the spherical Bessel function of order $\ell$,
and $g(\tau) = \dot \kappa e^{\kappa}$.
Physically, $g(\tau)$ is a visibility function, whose peak defines 
the epoch of recombination,
which gives the dominant contribution to observed anisotropies.
Note the double $g(\tau)$ integral in the polarization term; this
appears as during last scattering
the $T$ anisotropies are the source of $E$ polarization 
which in turn is the source of $B$ polarization.

By expressing $\cos \theta$ in terms of $Y_{1,0}$ and applying the three 
$Y_{\ell,m}$ formula, Eq. \ref{eq:crosscor} can be further simplified as
\begin{eqnarray}
\langle a_{\ell' m'}^{T*} a^B_{\ell m}\rangle  =     
C^{TB}_{\ell-1,\ell}(\hat{\bf{z}})
\left[\frac{\ell^2 - m^2}{4 \ell^2 - 1}\right]^{1/2} 
\delta_{\ell',\ell-1} \delta_{m,m'} +
C^{TB}_{\ell+1,\ell}(\hat{\bf{z}})
\left[\frac{(\ell+1)^2 - m^2}{4 (\ell + 1)^2 - 1}\right]^{1/2} 
\delta_{\ell',\ell+1} \delta_{m,m'}.
\label{eq:simplify}
\end{eqnarray}
The form of our result illustrates the SO(2) symmetry of the model 
which we are considering \cite{pedro}.  To see why this is so, consider the 
general case of $a_{\ell m}$'s that are generated by a 
model that is cylindrically symmetric
about the $z$ axis.  Then the symmetry transformations are
$R(\phi) a_{\ell m} = e^{i m \phi} a_{\ell m}$ and so  
$\langle a_{\ell' m'} a_{\ell m}\rangle = \delta_{mm'}C^{\ell\ell'}_{|m|}$
as terms with $m \neq m'$ are not invariant under $\phi$ rotations.
Furthermore, as we have found a nonzero correlation between coefficients of 
opposite parities, this indicates that the axis of symmetry has an 
orientation.  Our magnetic field lies along a definite axis and it points
a definite way along it. The effect of a ${\vec {\cal B}}_0$ pointing in 
the $x$ or $y$ directions can be similarly obtained by replacing the 
$\cos \theta$ in Eq.  \ref{eq:crosscor} with $x = \sin \theta \cos \phi$ and 
$y = \sin \theta \sin \phi$, yielding equations 
for $C^{TB}_{\ell\pm1,\ell}$ in the $x$ and $y$ directions.

\begin{figure}
\centerline{\psfig{file=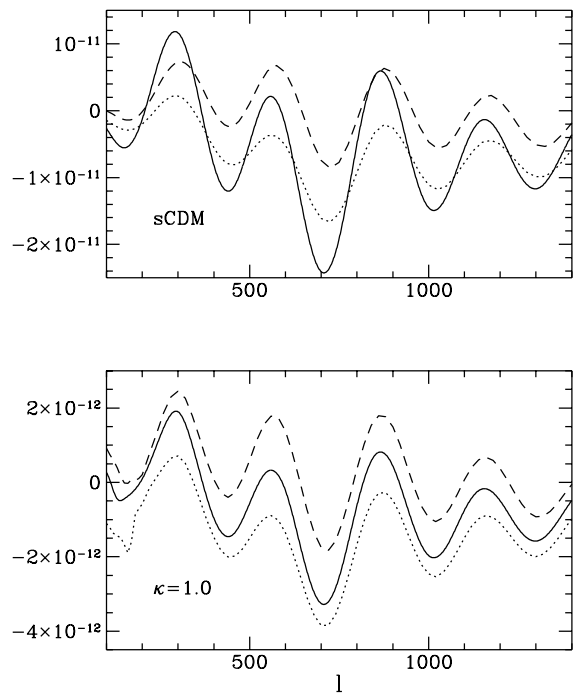,width=4in}}
\caption{ The top panel shows a 
comparison of $\ell(\ell+1)C^{TE}_{\ell,\ell}/(2\pi)$ (solid line),
$\ell(\ell+1)C^{TB}_{\ell-1,\ell}/(2\pi\omega_0)$ (dotted line)
and $\ell(\ell+1)C^{TB}_{\ell+1,\ell}/(2\pi\omega_0)$ (dashed line)
in a standard CDM cosmology, while the bottom panel shows
such a comparison in a reionized universe with $\kappa=1$.  }
\label {cls}
\end{figure}

Thus we have discovered measurable quantities that directly scale
with the magnitude and direction
of the magnetic field.  Experimentally, 
$C^{TB}_{\ell \pm1,\ell}(\hat{\bf{x_i}})$
can be determined by arbitrarily choosing a basis on the sky and then using
appropriate combinations of the $a^T$'s and $a^B$ to form estimators
in the $x$, $y$ and $z$ directions.   
We have calculated $C^{TB}_{\ell\pm1,l}$ using the CMBFAST code of
\cite{cmbfast}. We shall restrict ourselves to a standard cold dark
matter (CDM) universe to quantify the effect of the magnetic field; the
results appear in 
Figure \ref{cls} and where we normalize by the scale of the magnetic field
and plot $C^{TB}_{\ell\pm1,l}/\omega_0$. For comparison, we also
plot $C^{TE}_{\ell,\ell}$.  Note that if the last scattering surface were
taken to be infinitely thin, then $g(\tau)=\delta(\tau-\tau_*)$ and  
$C^{TB}_{\ell\pm1,\ell}/\omega_0$ and $C^{TE}_{\ell,\ell}$ would be identical
up to the difference introduced by integrating over two bessel functions
whose orders differ by one. 
As polarization is generated in the optically 
thin last scattering surface, the visibility function is 
quite sensitive to reionization. 
Thus we have also calcuated $C^{TB}_{\ell\pm1,\ell}$ for a 
CDM universe with an optical depth of $\kappa=1$ and we present
the results in the bottom panel of Figure \ref{cls}. Note that
reionization will supress power in the various cross correlations.
However the broader support of $g(\tau)$ leads to a slight increase
in amplitude of $C^{TB}_{\ell\pm1,l}/\omega_0$ relative to
$C^{TE}_{\ell,\ell}$.

\begin{figure}
\centerline{\psfig{file=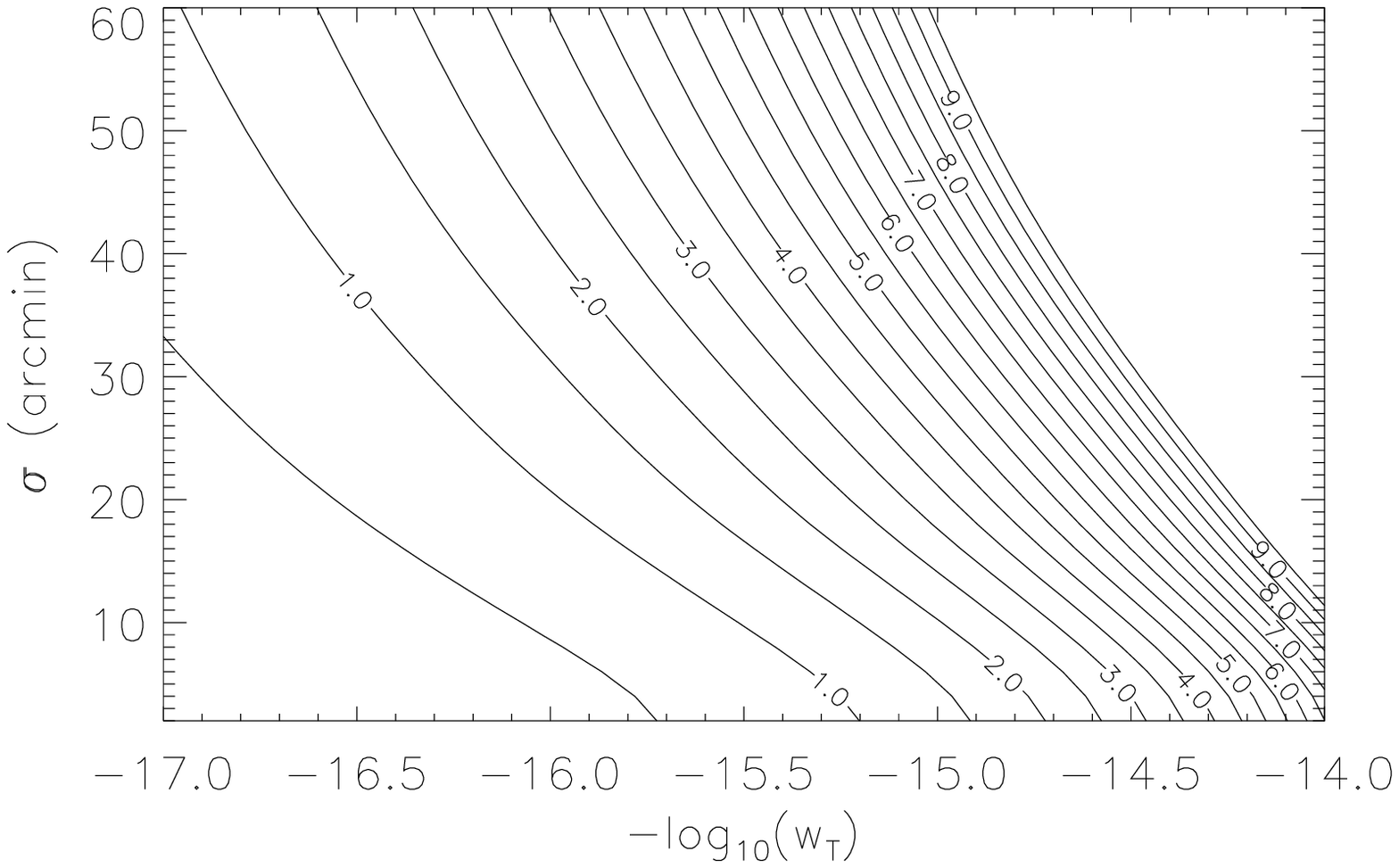,width=5.in}}
\centerline{\psfig{file=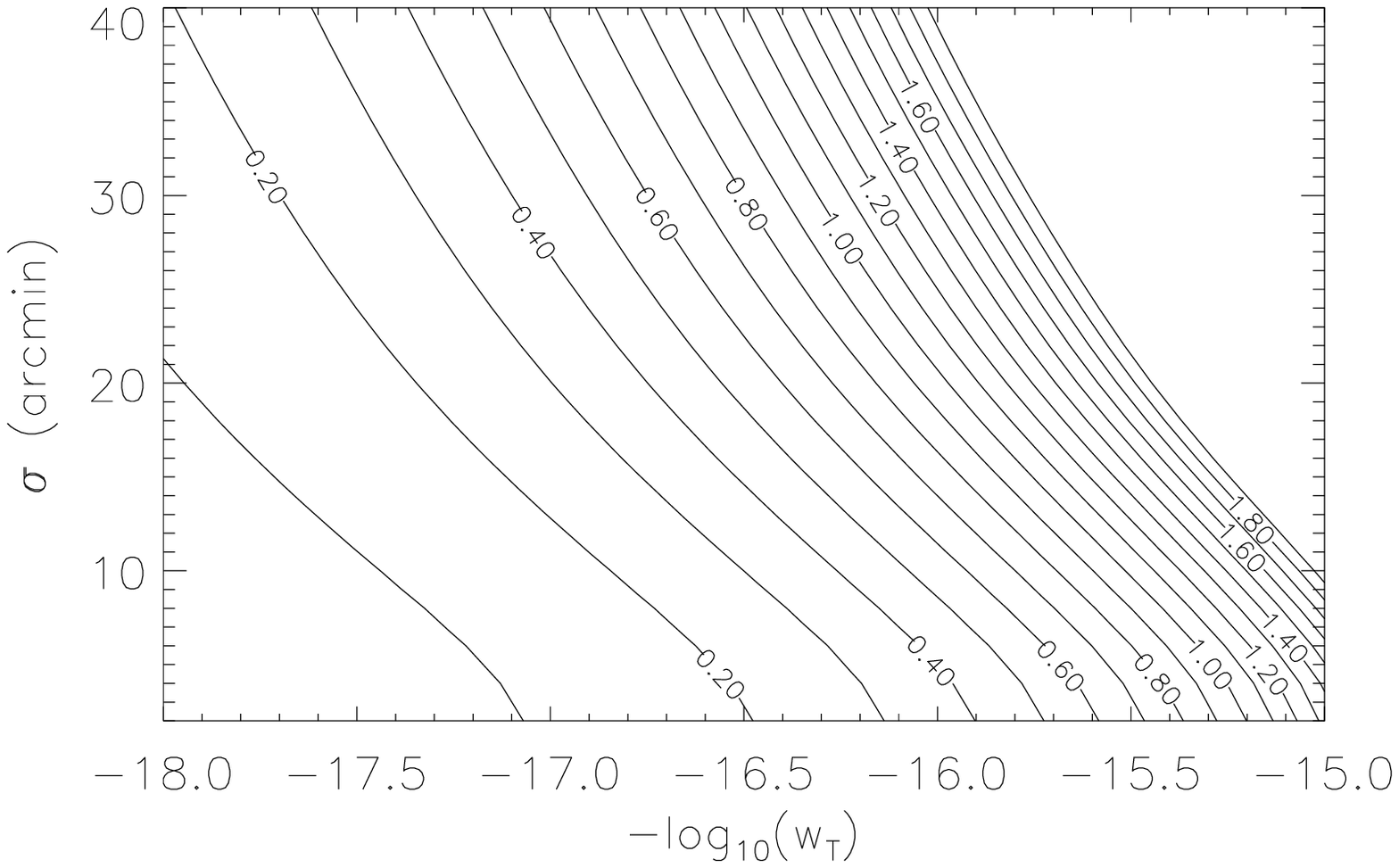,width=5.in}}
\caption{ The sensitivity of $(\frac{{\cal B}_0}{10^{-9}{\rm
G}})(\frac{\nu_0} {30{\rm GHz}})^{-2}$ given a full sky measurement
of the CMB temperature and anisotropy with noise characteristic
$w_T^{-1}$ and beamsize at full width half maximum $\sigma$ }
\label {exp}
\end{figure}

Let us now turn to the observability of such a signal. For the purpose
of this paper we shall consider a full sky survey such as would be
performed by a satellite.
To estimate the error in determining $C^{TB}$ we follow the analysis
given in \cite{h2}.  Being careful to include the $m$ coeficients in
Eq. \ref{eq:simplify} we find
\begin{eqnarray}
\Delta C^{TB}_{\ell\pm1,\ell} & \simeq  \frac{1}{2\ell} &\Big[ 2 \ln(2\ell)
\left( C^{TT}_{\ell+1} + w^{-1}_T e^{\ell^2 \sigma^2} \right)
\left(C^{BB}_{\ell} + w^{-1}_P e^{\ell^2 \sigma^2} \right) +
(C^{TB}_{\ell \pm1,\ell})^2 \Big]
\end{eqnarray}
where $\sigma^2$ is the full width at half maximum of the beam,
$w^{-1}_T$ and $w^{-1}_P$ are measures of the experimental
sensitivity, independent of pixel size and we have assumed
full sky coverage. We have also approximated 
\begin{eqnarray}
\sum_{m=-\ell}^{\ell}&\frac{1}{2\ell+1}&\left [\frac{4(\ell+1)^2-1}
{(\ell+1)^2-m^2}\right ]\nonumber \\  &\simeq& (2\ell+3)2\int_0^\ell \frac{dm}{(\ell+1)^2-m^2}
\end{eqnarray}
which is valid for large $\ell$'s.  Due to the
high cosmic variance and low signal at small $\ell$ we will want to
base our measurement of the magnetic field
on correlations with $\ell>50$ where this approximation is justified. 

Following \cite{jungman} we can define a function for the
goodness of fit of a theory given an observation:
\begin{eqnarray}
\chi^2(\omega_0)=\sum_{\ell}\left [
\frac{[C_{\ell+1,\ell}^{TBexp}-C_{\ell+1,\ell}^{TBth}(\omega_0)]}{(\Delta C^{TB}_{\ell+1,\ell})^2}+
\frac{[C_{\ell-1,\ell}^{TBexp}-C_{\ell-1,\ell}^{TBth}(\omega_0)]}{(\Delta C^{TB}_{\ell-1,\ell})^2}\right ]
\end{eqnarray}
where the sum is taken from $\ell > 50$ and
we assume that we know all other cosmological parameters.
We hope the true value of the magnetic field will minimize the $\chi^2$
and its sensitivity to ${\cal B}_0$ will tell us how well we
can constrain it. We are interested in determining
the minimum value of ${\cal B}_0$ to which we are sensitive.  In this
limit we expect noise to dominate over $C^{TB}$ and $C^{BB}$ as they
scale as ${\cal B}_0$ and ${\cal B}_0^2$ respectively, so we can neglect
these terms in Eq. 11.

Following \cite{nr}, an approximate 1$-\sigma$ error is
given by
\begin{eqnarray}
\sigma^{-2}(\omega_0)&=&\left
[\frac{1}{2}\frac{\partial^2\chi^2}{\partial^2{\omega_0}}
 \mid_{\omega_0=0} \right] \nonumber \\
&\simeq&\sum_{\ell}\frac{l}{\ln 2\ell}
\frac{(C_{\ell+1,\ell}^{TB}/\omega_0)^2+
(C_{\ell-1,\ell}^{TB}/\omega_0)^2}{(C^{TT}_{\ell}+w_T^{-1})
e^{2\ell^2\sigma^2} w_P^{-1}}
\end{eqnarray}

In Figure \ref{exp} we show two contour plots for the
expected precision in $(\frac{{\cal B}_0}{10^{-9}{\rm
G}})(\frac{\nu_0} {30{\rm GHz}})^{-2}$ for a range of experimental
parameters.  Here we take $w_P = 2 w_T$ \cite{h2}.
For low frequency detectors ($\nu_0<50 {\rm Ghz}$),
the top plot shows that we can get comparable limits to that
of \cite{BFS} for $w_T<10^{-15}$. Such
frequencies are currently only accessible through HEMT detectors
where the sensitivity has yet to reach such high levels.
With low frequencies there is the additional problem of having to
consider large beamwidths. For example an optimistic, space based
$30$GHz receiver can achieve at most $15'$ resolution.
The planned {\it MAP}  mission \cite{map} will have a $30$GHz with
$\sigma=42'$ and $w_T^{-1}\simeq4.7\times10^{-15}$; 
the {\it Planck} mission \cite{planck} will have a $31$GHz detector with 
$\sigma=30'$ and slightly smaller $w_T$. This means
we will get a constraint of 
$\sigma({\cal B}_0)\simeq {\rm few} \times 10^{-8}$Gauss
from both of them.

For high frequency measurements one is faced with the $\nu_0^2$ term,
i.e. the higher the frequency the smaller the effect.
In the bottom plot of Fig \ref{exp} we have consider
a lower range of $w_T$ appropriate for these measurements, and 
labeled isocontours in steps of 0.1. 
The important thing to note is that current
bolometer technology is reaching these levels of sensitivity.
If we consider the example of the $150$GHz detector on the
proposed {\it Planck} mission, we find that it may be possible to 
get a constraint of $\sigma({\cal B}_0)\simeq 10^{-9}$,
an improvement over the limits set in \cite{BFS}. With
the rapid advances in both high frequency HEMT and
bolometer technology it is conceivable that one may
do even better (for a realistic assessment of future prospects
see \cite{GS}). 

In this letter we have presented a new method for constraining
a homogeneous, primordial, magnetic field. In describing the
technique and its potential sensitivity, we have restricted
ourselves to a full sky measurement of the CMB anisotropy and
polarization.
Clearly this is a simplification. A full analysis should
include different experiments with varying degrees of sky 
coverage \cite{HM}. We have also focused on scalar perturbations
in a CDM universe. The inclusion of gravity waves and vorticity
will generate a non-negligible $C^{BB}$ which may increase
the cosmic variance of our statistic and necessarily reduce
its sensitivity. Naturally the inclusion of global
anisotropy, the effects of foreground sources
and different thermal histories may also change our results.
For example, for the reionized universe considered in Figure
\ref{cls} the constraint is worse by a factor of two as compared
to the sCDM case. Again a full analysis should 
include a wide range of cosmological 
parameters and scenarios.

This method has two appealing features which we 
restate. Firstly all current scenarios of structure
formation assume statistical isotropy which necessarily
leads to a zero cross correlation between B type polarization
and temperature anisotropy. It is only through the existence
of magnetic field (or some form of global anisotropy) 
that such a cross correlation can be induced.
Secondly the constraint derived in \cite{BFS} was strongly
dominated by the large angle anisotropies. The authors
pointed out that a good measurement of the temperature anisotropies
already exists on these scales and future measurements will
do little to improve it. This is not the case of our
method. Clearly, the better the measurements become the better
the constraint of ${\cal B}_0$ will be.

{\it Acknowledgments}: 
We would like to thank John Arons, John Barrow, 
Andrew Cummings, Andrew Jaffe,
Shaul Hanany, Joe Silk, and George Smoot for helpful discussions.  
E.S. is supported by an NSF fellowship.  P.F.
was supported by the Center for Particle Astrophysics, an NSF Science and
Technology Center at U.C.~Berkeley, under Cooperative Agreement No. AST
9120005. We thank U. Seljak and M. Zaldarriaga for the use of CMBFAST.


\begin{references}
\bibitem{a}  P. Kronenberg, Rep. Prog. Phys. {\bf 57}, 325 (1994);
Rees, M. J.  Quat. J.R.A.S., {\bf 28}, 197 (1987).

\bibitem{b}  R. Pudritz and J. Silk, Ap. J. {\bf 342}, 659 (1989); 
R. Kulsrud, R. Cen, J. P. Ostriker, and D. Ryu, Ap. J. {\bf 480},
481 (1997).

\bibitem{f}  J. M. Cornwall 1997, {\tt hep-th/9704022};
R. M. Kulsrud, R. Cen, J. Ostriker, and D. Ryu, Ap. J. {\bf 480}, 481 (1997); 
K. Enqvist and P. Olensen, Phys. Lett. {\bf B329}, 195 (1994); 
A. D. Dolgov and J. Silk, Phys. Rev. D {\bf 47}, 3144 (1993); 
B. Ratra, Ap. J.{\it \ }{\bf 391}, L1(1992);
T. Vachaspati, Phys. Lett. {\bf B265}, 258 (1991);
J. Quashnock, A. Loeb and D.N. Spergel, Ap. J. {\bf 344}, L49 (1989); 
M. S. Turner and L. M. Widrow, Phys. Rev. D {\bf 30}, 2743 (1988);  
E. R. Harrison, Mon. Not. R. astron. Soc. {\bf 165}, 185 (1973);

\bibitem{c}  S. I. Vainshtein, E. N. Parker, and R, Rosner, Ap. J. {\bf 404},
773 (1993);, S. I. Vainshtein, and F. Cattaneo, Ap. J., {\bf 393} 
165 (1992).

\bibitem{d}  P. P. Kronberg, J. J. Perry and E. L. Zukowski, Ap. J. 
{\bf 387}, 528 (1992).

\bibitem{e}  A. M. Wolfe, K. Lanzetta and A. L. Oren, Ap. J. {\bf 388},
17 (1992).

\bibitem{thorne} K. S. Thorne, Ap. J. {\bf 148}, 51 (1967); A. G. Doroshkevich
Astrophysics {\bf 1}, 138 (1967); K. Jacobs, Ap. J. {\bf 155}, 379 (1969).

\bibitem{adams} J. Adams, U. Danielsson, D. Grasso and H. Rubinstein,
 Phys. Lett. B {\bf 388} 253 (1996)

\bibitem{BFS} J. Barrow, P. Ferreira, and J. Silk, Phys. 
Rev. Lett. {\bf 78}, 3610 (1997). 

\bibitem{milan} E. Milaneschi, and R. Fabbri,  A\&A {\bf 151} 7 (1985).

\bibitem{hhz}
D. D. Harari, J. D. Hayward, and M. Zaldarriaga, Phys. Rev. D. 
{\bf 55}, 1840 (1997). 

\bibitem{KL}  A. Kosowsky and A. Loeb, Ap. J. {\bf 469} 1 (1996);

\bibitem{h} 
M. Zaldarriaga, and U. Seljak, Phys. Rev. D. {\bf 55}, 1830 (1997).

\bibitem{h2}
M. Kamionkowski, A. Kosowsky and A. Stebbins, 
Phys. Rev. D. {\bf 55} 7368 (1997).
 

\bibitem{i}  A. Kosowsky Annal. Phys. {\bf 246}, 49 (1996); S. 
Chandrasekhar {\it Radiative Transfer} (Dover, New York, 1960) Chapter 1.

\bibitem{MB} C.P. Ma and E. Bertschinger  Ap. J. {\bf
455} 7 (1995)

\bibitem{BT} C. Tsagas and J. Barrow, {\tt gr-qc/9704015}



\bibitem{pedro} P. G. Ferreira and J. U. Magueijo, 
{\tt astro-ph/9704052} submitted to Phy. Rev. D. 

\bibitem{cmbfast} U. Seljack and M. Zaldarriaga, Ap. J. {\bf 469} 437
(1996).

\bibitem{jungman} G. Jungamn, M. Kamionkowski, A. Kosowsky and
D. Spergel Phys. Rev. D {\bf 54} 1332 (1996)

\bibitem{nr} W. Press, S. Teukolsky, W. Vetterling, B. Flannery 
{\it Numerical Recipes} CUP (1992)

\bibitem{map} MAP home page, URL http://map.gsfc.nasa.gov

\bibitem{planck} 
Planck Mission home page, 
URL http://astro.estec.esa.nl:80/SA-general/ Projects/Cobras/cobras.html.

\bibitem{GS}G. Smoot {\tt astro-ph/9705135}
\bibitem{HM} M. Hobson and J. Magueijo M.N.R.A.S. {\bf 283}, 1133 (1996)

\end{references}
\end{document}